\documentclass[11pt, a4paper]{article}
\usepackage[numbers,sort&compress]{natbib}

\usepackage[utf8]{inputenc}
\usepackage[T1]{fontenc}
\usepackage{amssymb, amsmath, mathrsfs}   
\usepackage{graphicx} 
\usepackage[margin=2cm]{geometry}
\usepackage[colorlinks=true, allcolors=blue]{hyperref}
\usepackage{mathtools}
\usepackage{enumitem}

\begin{document}

\begin{center}
{\LARGE\bf Post-Newtonian expansion of scale-dependent gravity}\\

\vspace{.5cm}
\textbf{\large Nicolas R. Bertini}$^1$ and \textbf{\large Marcos H. Novaes}$^2$

\vspace{.5cm}
$^1$Departamento de Física, ICE, Universidade Federal de Juiz de Fora, \\
Campus Universitário - Juiz de Fora, 36036-900, MG, Brazil \\

\vspace{.3cm}  
$^2$PPGCosmo, CCE, Universidade Federal do Espírito Santo (UFES), \\
Vitória - ES, 29075-910, Brazil

\end{center}

\vspace{.3cm}
\begin{abstract}
\noindent
Scale-dependent gravity is an extension of general relativity in which the Newton and cosmological constants may vary slightly with the energy scale due to remnant low-energy quantum effects. A fundamental feature of this approach is the scale-setting procedure. In a previous work, a covariant expression for the scale, consistent with conservation laws, was established. Here, we apply the full Will–Nordtvedt version of the parameterized post-Newtonian (PPN) formalism to this framework, assuming that the constants can be expressed as a power series in the scale. We find that a new potential arises at the first post-Newtonian order, which is absent from the standard PPN formalism. This potential modifies the definitions of pressure and internal energy but does not affect center-of-mass orbits and is, therefore, not constrained by Solar System tests.
\end{abstract}

\section{Introduction}
 
Due to the success in explaining astronomical data at the Solar System scale \cite{Will:2014kxa} and predicting qualitatively new gravitational phenomena, such as recently observed gravitational waves \cite{LIGOScientific:2016aoc, LIGOScientific:2018mvr}, the general relativity (GR) is the standard theory for describing gravitational interaction. However, applying it on very large or very small scales leads to problems, both in the theoretical and phenomenological sectors, such as the inability to fully explain the nature of the dark sector \cite{Weinberg:1988cp, Faraoni:2010pgm, Famaey:2011kh, DelPopolo:2013evi, Amendola:2015ksp}, tensions involving some cosmological parameters \cite{Abdalla:2022yfr}, the prediction of singularities and the non-renormalizability at the quantum level. 

These issues indicate that GR must be extended, at the very least. In this context, the renormalization group (RG) has been applied to gravity as a theoretical tool to address the problem of non-renormalizability, offering a framework capable of incorporating quantum effects into GR at different energy scales. It is well established that quantum field theory (QFT) predicts, in general, distinct behaviors depending on the energy scale considered. Parameters such as coupling constants and effective masses can vary with the energy scale at which they are measured, and this dependence is characterized by the RG $\beta$-functions. In quantum electrodynamics, for instance, it can be shown that the coupling approaches a constant value in the infrared limit \cite{Goncalves:2009sk, KLOE-2:2016mgi}. 
In the case of gravity, the inclusion of higher-derivative terms in the Einstein–Hilbert action is required to ensure the renormalizability of the theory of quantum matter fields when coupled to classical gravity, or to enable a consistent quantization of the gravitational field itself. These terms are relevant at high energies but become dynamically negligible at large scales. Moreover, it can be shown that the running couplings associated with these terms have trivial RG flows in the infrared, meaning they asymptotically approach constant values \cite{Gorbar:2002pw, Gorbar:2003yt}.
On the other hand, the gravitational coupling (Newton's constant $G$) and the cosmological constant $\Lambda$ may exhibit a different behavior and vary even in the deep infrared regime (see, e.g., \cite{Shapiro:2008yu, Shapiro:2009dh}).  In this scenario, $G$ and $\Lambda$ are viewed as running couplings that depend on the renormalization scale $\mu$, and their scale dependence is determined by integrating the corresponding $\beta$-functions.

Although the fundamental gravitational action may include higher-derivative terms, such contributions are typically suppressed at low energies. Thus, GR is often viewed as an effective theory valid at certain energy scales. In this sense, since quantum corrections modify the couplings but may not introduce new dynamical degrees of freedom at all scales, the structure of Einstein’s equations can remain valid. Then the scale dependence of $G$ and $\Lambda$ represents small dynamical corrections that effectively capture RG effects and may be sufficient to account for phenomena at astrophysical and cosmological scales. However, the precise form of this scale dependence remains uncertain. The two key unknowns are the form of the physical $\beta$-functions in the low-energy regime --- which depends on the underlying QFT framework adopted, such as the effective field theory approach or asymptotic safety (see, e.g., \cite{Donoghue:1994dn, Reuter:1996cp, Bonanno:2000ep,  Giacchini:2020zrl}) --- and the identification of $\mu$ with physical quantities, i.e., the scale-setting procedure \cite{Reuter:2003ca, Shapiro:2004ch, Babic:2004ev, Domazet:2010bk, Koch:2014joa, Bertini:2024naa}.

In the work \cite{Bertini:2024naa}, a scale-setting procedure was developed that identifies $\mu$ with a scalar quantity naturally emerging from the analysis of the equations of motion. This identification reproduces the relations obtained via RG-based scale-setting approaches \cite{Shapiro:2004ch, Babic:2004ev, Domazet:2010bk, Koch:2014joa}, both in the cosmological context, where $\mu$ corresponds to the Hubble parameter, and in the astrophysical context with the source being a point-like mass, where $\mu\propto 1/r$, with $r$ the radial coordinate. Building on this framework, several physical systems can be proposed as candidates to investigate possible RG effects in gravity at large scales, and for any of them, Solar System bounds must be taken into account.

In this work, we apply the Will–Nordtvedt parametrized post-Newtonian (PPN) formalism \cite{poisson2014gravity, Will_2018}, to this scale-dependent framework. The PPN formalism is one of the most robust mechanisms for confronting gravity theories with observational data from Solar System tests. In this approach, the metric and matter variables --- where matter is modeled as a perfect fluid --- are expanded in powers of $v/c$, with $v$ being the typical velocity of the system and $c$ the speed of light. Observable deviations from GR are encoded in a set of ten parameters constrained by experimental bounds \cite{Will:2014kxa}.

A crucial requirement for applying the PPN formalism to compare alternative theories of gravity with GR is the conservation of the energy-momentum tensor $T^{\alpha\beta}$, in the form $\nabla_{\alpha}T^{\alpha\beta}=0$. This condition is necessary to ensure consistency with well-established weak-field predictions, enabling a coherent definition of the gravitational potentials that structure the perturbative expansion of the metric. In GR, this conservation law follows from the Bianchi identities, ensuring compatibility between matter dynamics and spacetime geometry. In scale-dependent gravity, the conservation law $\nabla_{\alpha}T^{\alpha\beta}=0$ together with the Bianchi identities, leads to a consistency relation that must be satisfied by $G(\mu)$ and $\Lambda(\mu)$. As a result, the running of these parameters cannot be specified independently: one of them is determined by renormalization group equations, while the other follows from the consistency relation. Here, we follow a common approach in the literature and assume that $G$ can be written as a power series in $\mu$ ( see, e.g., \cite{Shapiro:2004ch}, and references therein).

This paper is organized as follows. In the next section, we present the field equations, review the scale-setting procedure developed in \cite{Bertini:2024naa}, and establish the running laws. In Sec.~\ref{sec:PPN}, we perform the post-Newtonian expansion and discuss the results. Our conclusions and perspectives are given in Sec.~\ref{sec:conclusions}. In Appendix~\ref{EM-conservation}, we demonstrate that the covariant scale identification, together with the consistency relation, is sufficient to ensure the conservation of the energy-momentum tensor and the Appendix~\ref{PPN-exp} contains detailed calculations of the post-Newtonian expansion. We adopt the following notations and conventions: the Riemann tensor is defined as $R^{\alpha}_{\,\,\, \beta\sigma\rho} = \partial_{\sigma}\Gamma^{\alpha}_{\beta\rho} - \partial_{\rho}\Gamma^{\alpha}_{\beta\sigma}+\Gamma^{\alpha}_{\lambda\sigma}\Gamma^{\lambda}_{\beta\rho}- \Gamma^{\alpha}_{\lambda\rho}\Gamma^{\lambda}_{\beta\sigma}$, the Ricci tensor as $R_{\beta\rho}=R^{\alpha}_{\, \,\,\beta\alpha\rho}$ and the Ricci scalar as $R=g^{\alpha\beta}R_{\alpha\beta}$. The metric signature is taken to be $(-,+,+,+)$.

\section{General relativity with scale-dependent parameters}

Based on the arguments presented in the introduction, considering that the parameters $G$ and $\Lambda$ can vary according to a certain energy scale represented by $\mu$, the modified Einstein equations are
\begin{align}
G_{\alpha\beta} = 8\pi G(\mu) \big[ T_{\alpha\beta} - g_{\alpha\beta}\rho_{\Lambda}(\mu) \big] \,,\label{eq:FE}
\end{align}
where $G_{\alpha\beta} = R_{\alpha\beta} -g_{\alpha\beta}R/2$ is the Einstein tensor, $\rho_{\Lambda}\equiv \Lambda/8\pi G$ and $T_{\alpha\beta}$ is the energy-momentum tensor of the matter fields. Let us assume that $T_{\alpha\beta}$ represents a perfect fluid, i.e.
\begin{align}
T_{\alpha\beta} = (\varepsilon+p)u_{\alpha}u_{\beta} + g_{\alpha\beta}p\,, \label{eq:EMT}
\end{align}
where $p$, $\varepsilon$ and $u^{\alpha}$ are the pressure, energy density and the 4-velocity of the fluid, respectively.

A scale-setting procedure that is compatible with this scenario consists establishing a relationship between $\mu$ and the physical information contained in $\varepsilon$ \cite{Babic:2004ev,Shapiro:2004ch,Domazet:2010bk}. To this end, we adopt the approach developed in \cite{Bertini:2024naa}. By contracting Eq.~\eqref{eq:FE} with $u^{\alpha}u^{\beta}$ and using the energy-momentum tensor \eqref{eq:EMT}, we obtain
\begin{align}
    \varepsilon + \rho_\Lambda = \frac{1}{8\pi G}u^{\alpha}u^{\beta}G_{\alpha\beta}.\label{3}
\end{align}
In this relationship, the parameters $\rho_\Lambda$ and $G$ depend on $\mu$ and can be derived from some fundamental theory, such as the RG $\beta$-functions. Therefore, regardless of the specific forms for $G(\mu)$ and $\rho_{\Lambda}(\mu)$, the connection between $\mu$ and $\varepsilon$ is naturally expected to involve $u^{\alpha}u^{\beta}G_{\alpha\beta}$. In addition, Eq.~\eqref{3} shows that the projection $u^{\alpha}u^{\beta}G_{\alpha\beta}$ directly controls the local energy balance of the system. It is therefore natural to identify the renormalization scale with this same scalar quantity. This motivates the definition
\begin{align}
\mu^2 := \frac{1}{3}u^\alpha u^\beta G_{\alpha\beta}\,. 
\label{eq:scale-setting}
\end{align}
It is important to emphasize that this identification is not arbitrary. When matter is described as a perfect fluid, this scalar is the most natural covariant quantity constructed from the gravitational field equations that directly probes the local energy content measured by comoving observers. This choice is compatible with the requirement of energy-momentum conservation (see Appendix~\ref{EM-conservation}). Moreover, this identification reproduces the standard scale-setting relations obtained in previous renormalization-group–based approaches, yielding $\mu \sim H$, where $H$ is the Hubble parameter, in homogeneous and isotropic cosmology and $\mu \propto 1/r$ for static, spherically symmetric configurations. In this sense, Eq.~\eqref{eq:scale-setting} provides a minimal prescription for implementing the covariant scale dependence.

As demonstrated in Appendix ~\ref{EM-conservation}, the conservation law $\nabla_{\alpha}T^{\alpha\beta}=0$ holds with \eqref{eq:scale-setting} and provided that
\begin{align}
\frac{d\rho_\Lambda}{d\mu} - \frac{3\mu^2}{8\pi}\frac{dG^{-1}}{d\mu}=0\, \label{eq:consist}
\end{align}
along with the assumption of a fluid without vorticity\footnote{ Although this assumption is somewhat restrictive, it does not alter the structure of the post-Newtonian expansion, which remains organized in terms of the usual matter potentials.}.
This implies that the functional forms of $\rho_\Lambda$ and $G$ cannot be provided arbitrarily but must satisfy the above relation.

\subsection{Running couplings}

Typically, the running of one of the parameters is derived from the GR equations via the corresponding $\beta$-function,\footnote{The beta function of a coupling $X$ is defined as $\beta_X \equiv \mu\, \partial X / \partial \mu$.} while the other is obtained by solving Eq.~\eqref{eq:consist}. Several approaches implement a constant infrared $\beta$-function for $G^{-1}$ of the form $\beta_{G^{-1}} = 2\nu M_\text{Planck}^{2} = 2\nu G_{0}^{-1}$, where $\nu$ is a small dimensionless parameter. This leads to the logarithmic dependence $G^{-1} (\mu) = G_{0}^{-1}\big[1+\nu\ln(\mu^2/\mu_0^2)\big]$, where $\mu_0$ is the reference scale at which $G(\mu_0)=G_{0}$. The relation \eqref{eq:consist} then implies $\rho_{\Lambda}(\mu) = \rho_{\Lambda 0}+\frac{3\nu}{8\pi G_0}(\mu^2-\mu_0^2)$,where $\rho_{\Lambda 0 } = \rho_{\Lambda}(\mu_0)$. For the purposes of this paper, it is convenient to associate the RG scale with a scalar field $\varphi$ through the relation $\varphi = \mu^{2}$.

Following a similar approach to that of Ref.~\cite{Toniato:2017wmk}, we assume that $G^{-1}$ can be expanded as a function of $\varphi$ as
\begin{align}
G^{-1}=G_0^{-1} +\sum_n \nu_n \varphi^{n}\,.
\end{align}
Integrating Eq.~\eqref{eq:consist}, we obtain the corresponding expansion for $\rho_{\Lambda}(\varphi)$, which reads
\begin{align}
 \rho_{\Lambda} = \rho_{\Lambda 0}+  \frac{3}{8\pi}\sum_n \frac{n \nu_n}{n+1}\varphi^{n+1}\,, \label{eq:GandRho}
\end{align}
where the $\nu_n$ are real constants, and $G_0$ and $\rho_{\Lambda 0}$ denote the values of $G(\varphi)$ and $\rho_{\Lambda}(\varphi)$ at $\varphi = 0$. These expansions are not in general equivalent to the set of couplings $(G, \rho_\Lambda)$ in which $G$ exhibits a logarithmic dependence, but they can capture effects compatible with those arising when $\mu$ slightly deviates from $\mu_0$. Moreover, they are more suitable for applications within the PPN formalism.

With the above parameterization, the standard GR is recovered with $\nu_n=0$, in which case $\rho_{\Lambda 0}$ includes the usual cosmological constant. This constant leads to spacetimes that are not asymptotically flat and is therefore excluded from the standard PPN analysis. Moreover, given its value inferred from cosmological observations, its impact on Solar System dynamics is negligible (see e.g.~\cite{Sereno:2006re}). In fact, for a consistent application of the PPN formalism $\rho_{\Lambda 0}=0$ (see Sec.~\ref{sec:PPN} for details). On the other hand, to preserve the conservation law and the scale dependence discussed above, the expansion of the second term in $\rho_{\Lambda}(\varphi)$ must be retained; otherwise, either $\nabla_{\alpha}T^{\alpha\beta}\neq0$ or $G$ remains constant.

\section{Post-Newtonian expansion in scale-dependent gravity}\label{sec:PPN}

In this section we apply the Will-Nordvedt PPN formalism \cite{Will_2018} to the scale-dependent framework developed above. In this formalism, the source of the gravitational field is a perfect fluid and the main small parameter is the matter velocity field $|\vec{v}|=v\ll1$. The metric is expanded around the Minkowski spacetime, i.e.,
\begin{align}
    g_{\alpha\beta} = \eta_{\alpha\beta} + h_{\alpha\beta}\,,\label{eq:me-expand}
\end{align}
where $\eta_{\alpha\beta}$ is the Minkowski metric, which is of zeroth-order on $v$, and $h_{\alpha\beta}\sim {\cal O}(v^2)$, at least. For the first post-Newtonian order (1PN) one must determine $h_{00}$ up to $\mathcal{O}(v^4)$, $h_{0j}$ up to $\mathcal{O}(v^3)$, and $h_{ij}$ up to $\mathcal{O}(v^2)$.  This requires expanding the field equations \eqref{eq:FE} to the desired orders. For this purpose, it is convenient to rewrite them in the form
\begin{align}
	R_{\alpha\beta} = 8\pi G(\varphi) \left[ T_{\alpha\beta} -\frac{1}{2}g_{\alpha\beta}T + g_{\alpha\beta}\rho_{\Lambda}(\varphi) \right]\,.\label{eq:FE2}
\end{align}

Adopting the metric expansion \eqref{eq:me-expand} and taking into account that $u^{\alpha}=u^{0}(1,v^i)$, we have
\begin{align}
	\varphi = \frac{1}{3}\left[ (u^{0})^{2}(R_{00} + 2v^{i}R_{0i} + v^{i} v^{j} R_{ij}) +\frac{1}{2}(-R_{00} +\delta^{ij}R_{ij} -h^{00}R_{00} - 2 h^{0i}R_{0i} - h^{ij}R_{ij} ) \right]\,,\label{eq:phi-expand}
\end{align}
where $(u^{0})^{2} = (1- h_{00}-v^{2}-v^{i}v^{j}h_{ij}-2v^{i}h_{0i})^{-1}\,$. A consistent post-Newtonian expansion is achieved by considering $\varphi\sim{\cal O}(v^2)$ and $\sim{\cal O}(v^{4})$. The components of the Ricci tensor can be expressed as
\begin{align}
R_{00} =& -\frac{1}{2}\nabla^2h_{00} - \frac{1}{2}\left(h^k_{~k,00}- 2\,h^k_{~0,k0} \right) - \frac{1}{4}\,|\vec{\nabla}h_{00}|^2 +\frac{1}{2}\,h_{00,l}\left(h^{lk}_{~~,k}- \frac{1}{2}\,h^k_{~k,j}\delta^j_l\right)+ \frac{1}{2}\,h^{kl}h_{00,lk}\,,\end{align}
\begin{align}
R_{0i}=& -\frac{1}{2}\left(\nabla^2h_{0i} - h^k_{~0,ik} + h^k_{~k,0i} - h^k_{~i,k0} \right)\,,\label{2}
\end{align}
\begin{align}
R_{ij}=& -\frac{1}{2}\!\left(\nabla^2h_{ij} -\! h_{00,ij} +\! h^k_{~k,ij}-\! h^k_{~i,kj}- \! h^k_{~j,ki} \right),\label{eq:Rij}
\end{align}
where the comas refer to partial derivatives, $\nabla^2 \equiv \eta^{ij}\partial_i\partial_j$, and it is used that time derivatives effectively yield to a higher order in the expansion: if a quantity $X$ is of order $v^n$ then $X,_k\sim O(v^n)$ and $X,_0\sim O(v^{n+1})$.

Since the gravitational source is  a perfect fluid, $T_{\alpha\beta}$ is given by \eqref{eq:EMT} but the energy density is decomposed into the rest mass density $\rho$ and the specific internal energy $\Pi$ such that $\varepsilon = \rho+\rho\Pi$. Then
\begin{equation}
T^{\alpha\beta} = (\rho+\rho\Pi + p) (u^{0})^{2}\frac{dx^{\alpha}}{dt}\frac{dx^{\beta}}{dt} + (\eta^{\alpha\beta} - h^{\alpha\beta} + {\cal O}(h^2))p\,.\label{eq:EMT2}
\end{equation}
where $x^{\alpha} = (t,x^i)$ and $\rho\Pi\sim p\sim {\cal O}(v^4)$. 

Using the above relations together with the expression for $\varphi$ in \eqref{eq:phi-expand} we expand the Eq.~\eqref{eq:FE2} to compute the metric components order by order. As a first step, the zeroth-order equation trivially yields
\begin{align}
\rho_{\Lambda0}=0\,.
\end{align}
This result is expected, since such a constant would necessarily lead to non–asymptotically flat spacetimes --- just like the cosmological constant in standard GR --- which is not considered in the PPN framework.  This is also physically reasonable because, up to 1PN order and given its value as inferred from cosmological observations, its effect on local systems such as the Solar System is negligible \cite{Sereno:2006re}. 
A detailed derivation of the expansion of Eq.~\eqref{eq:FE2} to higher orders can be found in Appendix~\ref{PPN-exp}; in what follows, we present only the final results. From this point onward, we adopt the notation $\overset{(n)}{X}$  to indicate that $X\sim v^n$.  

For $\overset{(2)}h_{00}$ (the Newtonian limit), there are no corrections due to scale dependence, and agreement with local Newtonian gravity requires $G_0=G_N=1$. The same holds for the metric components $\overset{(2)}h_{ij}$ and $\overset{(3)}h_{0i}$. It then follows that, as in standard GR,
\begin{align}
    \overset{(2)}{h}_{00} = 2U, \qquad  \overset{(2)}{h}_{ij} = 2U\delta_{ij}, \qquad \overset{(3)}{h_{0j}}=-\frac{7}{2}V_{j}-\frac{1}{2}W_{j}\,.
\end{align}
Here $U$ is the Newtonian potential, defined by 
\begin{align}
    \nabla^{2}U = -4\pi \rho\,,\label{eq:poisson}
\end{align}
and the other potentials are defined by $\nabla^{2}V_i = -4\pi\rho v_j$ and $W_j = V_j - \chi,_{0j}$, where $\chi$ is an auxiliary potential satisfying $\nabla^{2}\chi = -2U$\,.

Corrections arise in the component $\overset{(4)}{h}_{00}$. Expanding Eq.~\eqref{eq:FE2} to the required order, we find that only the term $\overset{(2)}{\varphi}$ contributes. Using the previously obtained results, Eq.\eqref{eq:phi-expand} yields
\begin{align}
    \overset{(2)}{\varphi} = \frac{1}{6}\left[ \overset{(2)}{R}_{00}+\delta^{ij}\overset{(2)}{R}_{ij}\right] = -\frac{2}{3}\nabla^{2}U\,.
    \label{varphi}
\end{align}
Inserting this result into the expansion of Eq.~\eqref{eq:FE2}, we obtain the following expressions
\begin{align}
    g_{00} & = [g_{00}]_\text{GR}-\frac{32\pi\nu_1}{3}{\cal T},\label{eq:g00}
    \\
    g_{0i} & = [g_{0i}]_\text{GR},
    \\
    g_{ij} & = [g_{ij}]_\text{GR},\label{eq:gij}
\end{align}
where $[g_{\alpha\beta}]_\text{GR}$ is the GR post-Newtonian metric \cite{Will_2018} and ${\cal T}$ is a new potential defined as
\begin{align}
    \nabla^{2}{\cal T} = -4\pi \rho^2\,.
\end{align}

This potential was also found in the PN expansion of $f(R)$ Pallatini \cite{Toniato:2019rrd} and in conservative $f(R,T)$ gravity \cite{Bertini:2023pmp}. As discussed in \cite{Toniato:2019rrd}, this potential does not affect the center-of-mass motion of self-gravitating bodies at 1PN order. Its contribution can be absorbed into internal quantities such as pressure and internal energy, which are not directly observable in Solar System tests. Consequently, the standard PPN parameters remain unchanged, and Solar System observations cannot constrain the associated coupling.

It is worth noting that extensions of GR based on renormalization group predictions have been previously analyzed in the Solar System context \cite{Farina:2011me, Toniato:2017wmk}. In both cases, deviations from GR were found, allowing the free parameter of the model to be strongly constrained. The main difference compared to our results lies in the scale-setting procedure. In \cite{Farina:2011me}, the authors employed a method based on the dynamics of the Laplace–Runge–Lenz vector and identified $\mu \sim (\Phi_\text{Newt})^{\alpha}$, where $\Phi_\text{Newt}$ is the Newtonian potential and $\alpha$ is a free parameter. In that work, the gravitational source was assumed to be a point-like mass, for which $\Phi_\text{Newt} \propto 1/r$, coinciding with $\mu \propto 1/r$ obtained from \eqref{eq:scale-setting} when adopting the metric of a static and spherically symmetric spacetime \cite{Bertini:2024naa}. In contrast, here we consider the source to be a perfect fluid.

In Ref.~\cite{Toniato:2017wmk}, the PPN formalism was applied to an approach that introduces scale dependence at the action level (see \cite{Rodrigues:2015hba} for a review). This framework leads to equations of motion equivalent to scalar–tensor theories with a system-dependent potential. As in the previous case, the scale was identified with a function of the Newtonian potential which, even when perfect fluids are taken as sources, differs from the identification in \eqref{eq:scale-setting}, as shown in Eq.~\eqref{varphi}.
The difference between that approach and ours stems from two main aspects: the distinct scale-setting procedure adopted here, and the absence of derivatives of $G$ in our field equations. These features, taken together, suppress the post-Newtonian deviations reported in those earlier analyses.

Finally, we note that the validity of our result is unaffected if the running of $\rho_{\Lambda}$ is neglected; this only modifies the numerical factor multiplying ${\cal T}$ in Eq.~\eqref{eq:g00}.

\section{Conclusions}\label{sec:conclusions}

In this work, we have applied the full Will–Nordtvedt PPN to a covariant formulation of scale-dependent gravity. This framework incorporates a running gravitational coupling $G(\mu)$ and vacuum energy density $\rho_\Lambda(\mu)$ whose scale dependence is motivated by renormalization group arguments and fixed through a covariant scale-setting procedure. By ensuring energy-momentum conservation, the approach remains compatible with the foundational assumptions of the PPN formalism, enabling a systematic analysis of its weak-field predictions.

Our results show that, despite the modifications introduced by the scale dependence, the Newtonian and 1PN orders of the metric coincide with those of GR. The only deviation appears in the form of a new post-Newtonian potential proportional to $\rho^2$, which does not affect observable quantities in standard Solar System tests. Consequently, the PPN parameters take their GR values, and the theory remains consistent with current high-precision constraints.
This is a positive and nontrivial result, as the PPN formalism encapsulates some of the most accurately tested predictions of GR. The fact that the scale-dependent framework reproduces these results reinforces its viability. Furthermore, recent studies based on quantum field theory in curved spacetime indicate that classical tests such as the perihelion precession of planetary orbits remain in agreement with the known results \cite{Partanen:2025whp} (see also \cite{Sultana:2012qp} for similar results).

Our results reinforce the importance of investigating higher-order corrections in the post-Newtonian expansion, where deviations from general relativity may become observable (see \cite{Will:2025hwd} for effects related to the internal structure of compact bodies at the 3PN order). In this context, the analysis of gravitational waves emitted by binary systems, through extensions of the PPN formalism to higher orders, emerges as a particularly promising avenue. These aspects will be explored in future work.

\section{Acknowledgements}

We thank Ilya Shapiro for comments on this work. NRB acknowledges the support from \textit{Conselho Nacional de Desenvolvimento Científico e Tecnológico} (CNPq -Brazil).  MHN thanks \textit{Fundação de Amparo à Pesquisa e Inovação do Espírito Santo} (FAPES, Brazil) for support.

\appendix

\section{Energy-momentum tensor conservation}\label{EM-conservation}

In this section, we show that relations \eqref{eq:scale-setting} and \eqref{eq:consist} are sufficient to ensure that $\nabla_{\alpha}T^{\alpha\beta} = 0$, assuming an irrotational fluid. To this end, we follow a standard construction.  Since $g_{\alpha\beta}u^{\alpha}u^{\beta}=-1$, any arbitrary vector $A^{\alpha}$ can be decomposed into components parallel and orthogonal to $u^{\alpha}$ as
$ A^{\beta} = (-u_\alpha A^{\alpha})u^{\beta} + P^{\beta}_{\;\;\alpha}A^\alpha$,
where $P^{\alpha\beta} := g^{\alpha\beta}+u^{\alpha}u^{\beta}$ is a projector orthogonal to $u^{\alpha}$, i.e., $P^{\beta}_{\,\,\,\alpha}u^{\alpha}=0$ and $P^{\alpha}_{\;\;\sigma}P^{\sigma}_{\;\;\beta} = P^{\alpha}_{\;\;\beta}$. With this, we can write
\begin{align}
    \nabla_{\alpha}T^{\alpha\beta} = -(u^\sigma \nabla_{\alpha}T^{\alpha}_{\,\,\,\sigma})u^{\beta} + P^{\beta}_{\,\,\,\sigma}\nabla_{\alpha}T^{\alpha\sigma}\,.
\end{align}
Therefore, $\nabla_{\alpha}T^{\alpha\beta} =0$ if and only if
\begin{align}
u^\sigma \nabla_{\alpha}T^{\alpha}_{\,\,\,\sigma}&=0\,,\label{eq:cont}\\
P^{\beta}_{\,\,\,\sigma}\nabla_{\alpha}T^{\alpha\sigma}&=0\,.\label{eq:euler}
\end{align}
These equations are the relativistic versions of the continuity and Euler equations, respectively. The strategy is to show that these identities hold in the context of the scale-dependent gravity considered here.

Taking the divergence of Eq.~\eqref{eq:FE}, the Bianchi identities directly imply that
\begin{align}
    \nabla_{\alpha}T^{\alpha}_{\;\;\beta} = \nabla_\beta \rho_\Lambda - (T^{\alpha}_{\;\;\beta} - \rho_\Lambda \delta^{\alpha}_{\beta})G^{-1}\nabla_{\alpha}G\,.\label{eq:divT}
\end{align}
Contracting with $u^{\beta}$, considering the energy-momentum tensor of a perfect fluid \eqref{eq:EMT} and taking into account that $\nabla_{\alpha}\rho_\Lambda = (\nabla_{\alpha}\mu)d\rho_\Lambda/d\mu$ and $\nabla_{\alpha}G = (\nabla_{\alpha}\mu)dG/d\mu$, we obtain
\begin{align}
    u^{\beta}\nabla_{\alpha}T^{\alpha}_{\;\;\beta} = \left( \frac{d\rho_\Lambda}{d\mu} + \frac{\varepsilon+\rho_\Lambda}{G}\frac{dG}{d\mu} \right) u^{\alpha}\nabla_{\alpha}\mu\,.
\end{align}
Using relation \eqref{eq:scale-setting}, we conclude that in order to obtain $u^{\beta}\nabla_{\alpha}T^{\alpha}_{\;\;\beta} = 0$ with $u^{\alpha}\nabla_{\alpha}\mu\neq0$, it must be that
\begin{align}
    \frac{d\rho_\Lambda}{d\mu} - \frac{3\mu^2}{8\pi}\frac{dG^{-1}}{d\mu}=0, \label{eq:consist1}
\end{align}
where $\mu^2$ is given in Eq. \eqref{3}.

Let us now consider the equality \eqref{eq:euler}. Contracting \eqref{eq:divT} with $P^{\sigma\beta}$ and using Einstein's equation, we obtain
\begin{align}
    P^{\beta}_{\;\;\sigma}\nabla_{\alpha}T^{\alpha \sigma} = P^{\beta}_{\;\;\sigma}\left(\frac{d\rho_\Lambda}{d\mu}g^{\alpha\sigma} + \frac{G^{\alpha\sigma}}{8\pi}\frac{dG^{-1}}{d\mu} \right)\nabla_{\alpha}\mu\,.
\end{align}

Here is the key step that demonstrates the compatibility of the identification \eqref{3} with the conservation law. Using this identification along with \eqref{eq:consist1}, one gets
\begin{align}
    \nabla_{\alpha}\mu = \frac{1}{6\mu}\nabla_{\alpha}(u^{\sigma}u^{\rho}G_{\sigma\rho}) = \frac{4\pi}{3\mu}\nabla_{\alpha}[G(\varepsilon+\rho_\Lambda)] = \frac{4\pi G}{3\mu}\nabla_{\alpha}\varepsilon\,.
\end{align}
Thus,
\begin{align}
    P^{\beta}_{\;\;\sigma}\nabla_{\alpha}T^{\alpha \sigma} = P^{\beta}_{\;\;\sigma}\big(3\mu^{2} g^{\alpha\sigma} + G^{\alpha\sigma}\big)\frac{G}{6\mu}\frac{dG^{-1}}{d\mu} \nabla_{\alpha}\varepsilon\,.\label{eq:PnablaT}
\end{align}

For the right-hand side of this equation to vanish, it is necessary that $\big(3\mu^{2} g^{\alpha\sigma} + G^{\alpha\sigma}\big)\nabla_{\alpha}\varepsilon \propto u^{\sigma}$. One way to ensure this proportionality is to consider a barotropic fluid without vorticity, i.e., $\omega_{\alpha\beta}=P_{\alpha}^{\;\mu}P_{\beta}^{\;\nu}\nabla_{[\mu}u_{\nu]}=0$. In this case, there exists locally a scalar potential $\theta(x^{\alpha})$ and a function $F(x^{\alpha})$ such that \cite{1967rhm..book.....L}
\begin{align}
    u_\alpha = F\nabla_\alpha\theta\,,
\end{align}
with $F = (-\nabla_{\alpha}\theta \nabla^{\alpha}\theta)^{-1/2}$ fixed by the normalization condition $u^{\alpha}u_{\alpha}=-1$. In the barotropic case, the condition $\omega_{\alpha\beta}=0$ is preserved along the dynamical evolution \cite{landau_fluid_mechanics, Anile_1990}. Moreover, the continuity equation $u^{\sigma}\nabla_{\alpha}T^{\alpha}_{\;\;\sigma} = -u^{\alpha}\nabla_{\alpha}\varepsilon - \nabla_{\alpha}u^{\alpha}(\varepsilon+p)=0$ implies that $\varepsilon = \varepsilon(\theta)$. Consequently,
\begin{align}
    \nabla_{\alpha}\varepsilon = \varepsilon' \nabla_{\alpha}\theta = \varepsilon' F^{-1}u_{\alpha}\,,
\end{align}
where $\varepsilon' = d\varepsilon/d\theta$. Substituting this relation into \eqref{eq:PnablaT} we obtain
\begin{align}
    P^{\beta}_{\;\;\sigma}\nabla_{\alpha}T^{\alpha \sigma} = P^{\beta}_{\;\;\sigma}\big(3\mu^{2} u^{\sigma} + G^{\alpha\sigma}u_{\alpha}\big)\frac{G\varepsilon'}{6\mu F}\frac{dG^{-1}}{d\mu}\,.
\end{align}
Since $P^{\beta}_{\;\;\sigma}u^{\sigma}=0$, the first term in parentheses vanishes automatically. The second term can be rewritten as
\begin{align}
    P^{\beta}_{\;\;\sigma}G^{\alpha\sigma}u_{\alpha} = (\delta^{\beta}_{\sigma}+u^{\beta}u_{\sigma})G^{\alpha\sigma}u_{\alpha} = u_\alpha G^{\alpha\beta}+ 3u^{\beta}\mu^{2}\,,
\end{align}
which vanishes only if
\begin{align}
    u_\alpha G^{\alpha\beta} = -3u^{\beta}\mu^{2}.
\end{align}

This condition is indeed satisfied, as both sides reduce to $3\mu^2$ when contracted with $u^\beta$, consistently with the definition in \eqref{eq:scale-setting}.

\section{Post-Newtonian expansion}\label{PPN-exp}

In this appendix, we carry out the post-Newtonian expansion of scale-dependent gravity using the PPN formalism. To this end,  we impose the following set of gauge conditions
\begin{align}
    \partial_\alpha h^\alpha_i - \frac{1}{2} \partial_i h^\alpha_\alpha &= 0, \label{eq:gauge_spatial_appendix}  \\ % Usei A1, A2 etc. para o apêndice
    \partial_\alpha h^\alpha_0 - \frac{1}{2} \partial_0 h^\alpha_\alpha &= -\frac{1}{2}\partial_0 h_{00}. \label{eq:gauge_time_appendix}
\end{align}
Thus, using the notation established in Sec.~\ref{sec:PPN}, the Ricci tensor components expanded to the required order can be expressed as \cite{Will_2018}
\begin{align}
    R_{00} &= -\frac{1}{2} \nabla^2 \overset{(2)}{h}_{00} -\frac{1}{2} \nabla^2 \overset{(4)}{h}_{00} - \frac{1}{2} \Big(\partial_i \overset{(2)}{h}_{00}\Big)^2 + \frac{1}{2} \overset{(2)}{h}{}^{ij} \partial_i \partial_j \overset{(2)}{h}_{00} + \mathcal{O}(v^6), \label{R00_appendix}  \\
    R_{0j} &= -\frac{1}{2} \nabla^2 \overset{(3)}{h}_{0j} - \frac{1}{4} \partial_j \partial_0 \overset{(2)}{h}_{00} + \mathcal{O}(v^5), \label{R0j_appendix} \\
    R_{ij} &= -\frac{1}{2} \nabla^2 \overset{(2)}{h}_{ij} + \mathcal{O}(v^4). \label{Rij_appendix} 
\end{align}

For the second order (the Newtonian limit), the Eq.~\eqref{eq:FE2} is
\begin{align}
	\overset{(2)}{R_{00}} = 8\pi G_{0}\left[\overset{(2)}{ T_{00}} -\frac{1}{2}\eta_{00}\overset{(2)}{T} \right]\,.
	\quad
\end{align}
The energy-momentum tensor \eqref{eq:EMT2} provides $\overset{(2)}{T}_{00}=\rho$ and $\overset{(2)}T=-\rho$. Then, using Eq.~\eqref{R00_appendix} restricted to this order and the Poisson equation \eqref{eq:poisson}, we find
\begin{align}
    \overset{(2)}{h_{00}} = 2G_0 U\,,\label{eq:h002}
\end{align}
then the identification $G_0 = G_N = 1$ ensures the agreement with the local Newtonian gravity.

The field equations for the spatial components has analogous derivation. Within the same perturbative order
\begin{equation}
    \overset{(2)}{R_{ij}}=8\pi \left[\overset{(2)}{T_{ij}}-\frac{1}{2}\eta_{ij}\overset{(2)}{T}\right].
\end{equation}
Using Eq.~\eqref{Rij_appendix}, and noting that $\overset{(2)}T_{ij} = 0$, one gets
\begin{align}
    \overset{(2)}{h_{ij}}=2U\delta_{ij}\,.
\end{align}
Using this result together with \eqref{eq:h002}, we calculate the corresponding order of $\varphi$ from \eqref{eq:phi-expand} and find the relationship in \eqref{varphi}. 
As in the cases above, for $\overset{(3)}{h}_{ij}$ the theory introduces no deviations from standard GR, and thus the metric perturbation  can be written as 
\begin{align}
    \overset{(3)}{h_{0j}}=-\frac{7}{2}V_{j}-\frac{1}{2}W_{j}\,.
\end{align}

To calculate the metric component $\overset{(4)}{h_{00}}$, we expand the field equation \eqref{eq:FE} to the required order, i.e,
\begin{align}
	\overset{(4)}{R}_{00} = 8\pi \overset{(2)}{G}\left[ \overset{(2)}{T}_{00} - \frac{1}{2}\eta_{00}\overset{(2)}{T} \right] + 8\pi \left[ \overset{(4)}{T}_{00}-\frac{1}{2}\eta_{00}\overset{(4)}{T} - \frac{1}{2}\overset{(2)}{h}_{00}\overset{(2)}{T} + \eta_{00}\overset{(4)}{\rho_\Lambda} \right]\,. \label{eq:R00_v4} 
\end{align}
A straightforward expansion of the scale-dependent parameters reveals that
\begin{align}
    \overset{(2)}{G} = -\nu_1 \overset{(2)}{\varphi},\qquad \overset{(4)}{\rho}_{\Lambda}=\frac{3\nu_1}{16\pi}\Big[ \overset{(2)}{\varphi} \Big]^2\,.
\end{align}
Then, using \eqref{eq:phi-expand} and the Poisson equation, we get
\begin{align}
    \overset{(4)}{R}_{00}  = \overset{(4)}{R}{}^\text{GR}_{00}  - \frac{64}{3}\pi^{2}\nu_1 \rho^2\,,\label{eq:R00_o4}
\end{align}
with
\begin{equation}
\overset{(4)}{R}{}^{GR}_{00} = 8\pi \left( \overset{(4)}{T}_{00} - \frac{1}{2}\eta_{00} \overset{(4)}{T} - \frac{1}{2}\overset{(2)}{h}_{00} \overset{(2)}{T} \right) = \nabla^{2}(-2\phi_1 +2\phi_2 - \phi_3 - 3\phi_4),
\label{eq:R00_4RG}
\end{equation}
where we introduce the potentials
\begin{align}
    \nabla^{2}\phi_1 = -4\pi \rho v^2, \quad \nabla^{2}\phi_2 = -4\pi \rho U, \quad \nabla^{2}\phi_3 = -4\pi \rho\Pi, \quad \nabla^{2}\phi_4 = -4\pi p\,.
\end{align}
These are the terms obtained in standard GR. Using the definition of $\phi_2$, one can write
\begin{align}
    \overset{(4)}{R}_{00} = -\frac{1}{2}\nabla^{2}(\overset{(2)}{h}_{00}+2U^2-8\phi_2).
\end{align}
Then, integrating the field equation \eqref{eq:R00_o4} yields the standard time-time component of the expanded GR metric, which we denote by $\overset{(4)}{[h_{00}]}_\text{GR} = -2U^{2}+4\phi_1 + 4\phi_2+2\phi_3+6\phi_4$, plus corrections provided by scale dependence, i.e.,
\begin{align}
    \overset{(4)}{h_{00}} & = [\overset{(4)}{h}_{00}]_\text{GR}-\frac{32\pi\nu_1}{3}{\cal T},
\end{align}
where we define the new potential
\begin{align}
    {\cal T}(t,\vec{x}) \equiv \int \frac{[\rho(t,\vec{x}')]^{2}}{|x-\vec{x}'|}d^{3}x', \quad       \nabla^{2}{\cal T} = -4\pi\rho^{2}\,.
\end{align}
Therefore, the results can be summarized in Eqs.~\eqref{eq:g00}-\eqref{eq:gij}

\bibliographystyle{apsrev4-1}
%merlin.mbs apsrev4-1.bst 2010-07-25 4.21a (PWD, AO, DPC) hacked
%Control: key (0)
%Control: author (72) initials jnrlst
%Control: editor formatted (1) identically to author
%Control: production of article title (-1) disabled
%Control: page (0) single
%Control: year (1) truncated
%Control: production of eprint (0) enabled
%

%\bibliography{Nbib.bib}

\begin{thebibliography}{39}%
\makeatletter
\providecommand \@ifxundefined [1]{%
 \@ifx{#1\undefined}
}%
\providecommand \@ifnum [1]{%
 \ifnum #1\expandafter \@firstoftwo
 \else \expandafter \@secondoftwo
 \fi
}%
\providecommand \@ifx [1]{%
 \ifx #1\expandafter \@firstoftwo
 \else \expandafter \@secondoftwo
 \fi
}%
\providecommand \natexlab [1]{#1}%
\providecommand \enquote  [1]{``#1''}%
\providecommand \bibnamefont  [1]{#1}%
\providecommand \bibfnamefont [1]{#1}%
\providecommand \citenamefont [1]{#1}%
\providecommand \href@noop [0]{\@secondoftwo}%
\providecommand \href [0]{\begingroup \@sanitize@url \@href}%
\providecommand \@href[1]{\@@startlink{#1}\@@href}%
\providecommand \@@href[1]{\endgroup#1\@@endlink}%
\providecommand \@sanitize@url [0]{\catcode `\\12\catcode `\$12\catcode
  `\&12\catcode `\#12\catcode `\^12\catcode `\_12\catcode `\%12\relax}%
\providecommand \@@startlink[1]{}%
\providecommand \@@endlink[0]{}%
\providecommand \url  [0]{\begingroup\@sanitize@url \@url }%
\providecommand \@url [1]{\endgroup\@href {#1}{\urlprefix }}%
\providecommand \urlprefix  [0]{URL }%
\providecommand \Eprint [0]{\href }%
\providecommand \doibase [0]{http://dx.doi.org/}%
\providecommand \selectlanguage [0]{\@gobble}%
\providecommand \bibinfo  [0]{\@secondoftwo}%
\providecommand \bibfield  [0]{\@secondoftwo}%
\providecommand \translation [1]{[#1]}%
\providecommand \BibitemOpen [0]{}%
\providecommand \bibitemStop [0]{}%
\providecommand \bibitemNoStop [0]{.\EOS\space}%
\providecommand \EOS [0]{\spacefactor3000\relax}%
\providecommand \BibitemShut  [1]{\csname bibitem#1\endcsname}%
\let\auto@bib@innerbib\@empty
%</preamble>
\bibitem [{\citenamefont {Will}(2014)}]{Will:2014kxa}%
  \BibitemOpen
  \bibfield  {author} {\bibinfo {author} {\bibfnamefont {C.~M.}\ \bibnamefont
  {Will}},\ }\href {\doibase 10.12942/lrr-2014-4} {\bibfield  {journal}
  {\bibinfo  {journal} {Living Rev. Rel.}\ }\textbf {\bibinfo {volume} {17}},\
  \bibinfo {pages} {4} (\bibinfo {year} {2014})},\ \Eprint
  {http://arxiv.org/abs/1403.7377} {arXiv:1403.7377 [gr-qc]} \BibitemShut
  {NoStop}%
%%CITATION = ARXIV:1403.7377;%%
\bibitem [{\citenamefont {Abbott}\ \emph {et~al.}(2016)\citenamefont {Abbott}
  \emph {et~al.}}]{LIGOScientific:2016aoc}%
  \BibitemOpen
  \bibfield  {author} {\bibinfo {author} {\bibfnamefont {B.~P.}\ \bibnamefont
  {Abbott}} \emph {et~al.} (\bibinfo {collaboration} {LIGO Scientific,
  Virgo}),\ }\href {\doibase 10.1103/PhysRevLett.116.061102} {\bibfield
  {journal} {\bibinfo  {journal} {Phys. Rev. Lett.}\ }\textbf {\bibinfo
  {volume} {116}},\ \bibinfo {pages} {061102} (\bibinfo {year} {2016})},\
  \Eprint {http://arxiv.org/abs/1602.03837} {arXiv:1602.03837 [gr-qc]}
  \BibitemShut {NoStop}%
%%CITATION = ARXIV:1602.03837;%%
\bibitem [{\citenamefont {Abbott}\ \emph {et~al.}(2019)\citenamefont {Abbott}
  \emph {et~al.}}]{LIGOScientific:2018mvr}%
  \BibitemOpen
  \bibfield  {author} {\bibinfo {author} {\bibfnamefont {B.~P.}\ \bibnamefont
  {Abbott}} \emph {et~al.} (\bibinfo {collaboration} {LIGO Scientific,
  Virgo}),\ }\href {\doibase 10.1103/PhysRevX.9.031040} {\bibfield  {journal}
  {\bibinfo  {journal} {Phys. Rev. X}\ }\textbf {\bibinfo {volume} {9}},\
  \bibinfo {pages} {031040} (\bibinfo {year} {2019})},\ \Eprint
  {http://arxiv.org/abs/1811.12907} {arXiv:1811.12907 [astro-ph.HE]}
  \BibitemShut {NoStop}%
\bibitem [{\citenamefont {Weinberg}(1989)}]{Weinberg:1988cp}%
  \BibitemOpen
  \bibfield  {author} {\bibinfo {author} {\bibfnamefont {S.}~\bibnamefont
  {Weinberg}},\ }\href {\doibase 10.1103/RevModPhys.61.1} {\bibfield  {journal}
  {\bibinfo  {journal} {Rev. Mod. Phys.}\ }\textbf {\bibinfo {volume} {61}},\
  \bibinfo {pages} {1} (\bibinfo {year} {1989})}\BibitemShut {NoStop}%
%%CITATION = RMPHA,61,1;%%
\bibitem [{\citenamefont {Faraoni}\ and\ \citenamefont
  {Capozziello}(2011)}]{Faraoni:2010pgm}%
  \BibitemOpen
  \bibfield  {author} {\bibinfo {author} {\bibfnamefont {V.}~\bibnamefont
  {Faraoni}}\ and\ \bibinfo {author} {\bibfnamefont {S.}~\bibnamefont
  {Capozziello}},\ }\href {\doibase 10.1007/978-94-007-0165-6} {\emph {\bibinfo
  {title} {{Beyond Einstein Gravity}: {A Survey of Gravitational Theories for
  Cosmology and Astrophysics}}}},\ \bibinfo {series} {Fundamental Theories of
  Physics}, Vol.\ \bibinfo {volume} {170. 170}\ (\bibinfo  {publisher}
  {Springer},\ \bibinfo {address} {Dordrecht},\ \bibinfo {year}
  {2011})\BibitemShut {NoStop}%
%%CITATION = FTPHD,170,;%%
\bibitem [{\citenamefont {Famaey}\ and\ \citenamefont
  {McGaugh}(2012)}]{Famaey:2011kh}%
  \BibitemOpen
  \bibfield  {author} {\bibinfo {author} {\bibfnamefont {B.}~\bibnamefont
  {Famaey}}\ and\ \bibinfo {author} {\bibfnamefont {S.}~\bibnamefont
  {McGaugh}},\ }\href {\doibase 10.12942/lrr-2012-10} {\bibfield  {journal}
  {\bibinfo  {journal} {Living Rev. Rel.}\ }\textbf {\bibinfo {volume} {15}},\
  \bibinfo {pages} {10} (\bibinfo {year} {2012})},\ \Eprint
  {http://arxiv.org/abs/1112.3960} {arXiv:1112.3960 [astro-ph.CO]} \BibitemShut
  {NoStop}%
%%CITATION = ARXIV:1112.3960;%%
\bibitem [{\citenamefont {Del~Popolo}(2013)}]{DelPopolo:2013evi}%
  \BibitemOpen
  \bibfield  {author} {\bibinfo {author} {\bibfnamefont {A.}~\bibnamefont
  {Del~Popolo}},\ }\bibfield  {booktitle} {\emph {\bibinfo {booktitle}
  {{Proceedings, 9th Mexican School on Gravitation and Mathematical Physics:
  Cosmology for the XXI Century: Inflation, Dark Matter and Dark Energy
  (DGFM-SMF): Puerto Vallarta, Jalisco, Mexico, December 3-7, 2012}}},\ }\href
  {\doibase 10.1063/1.4817029} {\bibfield  {journal} {\bibinfo  {journal} {AIP
  Conf. Proc.}\ }\textbf {\bibinfo {volume} {1548}},\ \bibinfo {pages} {2}
  (\bibinfo {year} {2013})}\BibitemShut {NoStop}%
%%CITATION = APCPC,1548,2;%%
\bibitem [{\citenamefont {Amendola}\ and\ \citenamefont
  {Tsujikawa}(2015)}]{Amendola:2015ksp}%
  \BibitemOpen
  \bibfield  {author} {\bibinfo {author} {\bibfnamefont {L.}~\bibnamefont
  {Amendola}}\ and\ \bibinfo {author} {\bibfnamefont {S.}~\bibnamefont
  {Tsujikawa}},\ }\href@noop {} {\emph {\bibinfo {title} {{Dark Energy}:
  {Theory and Observations}}}}\ (\bibinfo  {publisher} {Cambridge University
  Press},\ \bibinfo {year} {2015})\BibitemShut {NoStop}%
\bibitem [{\citenamefont {Abdalla}\ \emph {et~al.}(2022)\citenamefont {Abdalla}
  \emph {et~al.}}]{Abdalla:2022yfr}%
  \BibitemOpen
  \bibfield  {author} {\bibinfo {author} {\bibfnamefont {E.}~\bibnamefont
  {Abdalla}} \emph {et~al.},\ }\href {\doibase 10.1016/j.jheap.2022.04.002}
  {\bibfield  {journal} {\bibinfo  {journal} {JHEAp}\ }\textbf {\bibinfo
  {volume} {34}},\ \bibinfo {pages} {49} (\bibinfo {year} {2022})},\ \Eprint
  {http://arxiv.org/abs/2203.06142} {arXiv:2203.06142 [astro-ph.CO]}
  \BibitemShut {NoStop}%
\bibitem [{\citenamefont {Goncalves}\ \emph {et~al.}(2009)\citenamefont
  {Goncalves}, \citenamefont {de~Berredo-Peixoto},\ and\ \citenamefont
  {Shapiro}}]{Goncalves:2009sk}%
  \BibitemOpen
  \bibfield  {author} {\bibinfo {author} {\bibfnamefont {B.}~\bibnamefont
  {Goncalves}}, \bibinfo {author} {\bibfnamefont {G.}~\bibnamefont
  {de~Berredo-Peixoto}}, \ and\ \bibinfo {author} {\bibfnamefont {I.~L.}\
  \bibnamefont {Shapiro}},\ }\href {\doibase 10.1103/PhysRevD.80.104013}
  {\bibfield  {journal} {\bibinfo  {journal} {Phys. Rev. D}\ }\textbf {\bibinfo
  {volume} {80}},\ \bibinfo {pages} {104013} (\bibinfo {year} {2009})},\
  \Eprint {http://arxiv.org/abs/0906.3837} {arXiv:0906.3837 [hep-th]}
  \BibitemShut {NoStop}%
%%CITATION = 0906.3837;%%
\bibitem [{\citenamefont {Anastasi}\ \emph {et~al.}(2017)\citenamefont
  {Anastasi} \emph {et~al.}}]{KLOE-2:2016mgi}%
  \BibitemOpen
  \bibfield  {author} {\bibinfo {author} {\bibfnamefont {A.}~\bibnamefont
  {Anastasi}} \emph {et~al.} (\bibinfo {collaboration} {KLOE-2}),\ }\href
  {\doibase 10.1016/j.physletb.2016.12.016} {\bibfield  {journal} {\bibinfo
  {journal} {Phys. Lett. B}\ }\textbf {\bibinfo {volume} {767}},\ \bibinfo
  {pages} {485} (\bibinfo {year} {2017})},\ \Eprint
  {http://arxiv.org/abs/1609.06631} {arXiv:1609.06631 [hep-ex]} \BibitemShut
  {NoStop}%
\bibitem [{\citenamefont {Gorbar}\ and\ \citenamefont
  {Shapiro}(2003{\natexlab{a}})}]{Gorbar:2002pw}%
  \BibitemOpen
  \bibfield  {author} {\bibinfo {author} {\bibfnamefont {E.~V.}\ \bibnamefont
  {Gorbar}}\ and\ \bibinfo {author} {\bibfnamefont {I.~L.}\ \bibnamefont
  {Shapiro}},\ }\href {\doibase 10.1088/1126-6708/2003/02/021} {\bibfield
  {journal} {\bibinfo  {journal} {JHEP}\ }\textbf {\bibinfo {volume} {02}},\
  \bibinfo {pages} {021} (\bibinfo {year} {2003}{\natexlab{a}})},\ \Eprint
  {http://arxiv.org/abs/hep-ph/0210388} {arXiv:hep-ph/0210388} \BibitemShut
  {NoStop}%
%%CITATION = HEP-PH/0210388;%%
\bibitem [{\citenamefont {Gorbar}\ and\ \citenamefont
  {Shapiro}(2003{\natexlab{b}})}]{Gorbar:2003yt}%
  \BibitemOpen
  \bibfield  {author} {\bibinfo {author} {\bibfnamefont {E.~V.}\ \bibnamefont
  {Gorbar}}\ and\ \bibinfo {author} {\bibfnamefont {I.~L.}\ \bibnamefont
  {Shapiro}},\ }\href {\doibase 10.1088/1126-6708/2003/06/004} {\bibfield
  {journal} {\bibinfo  {journal} {JHEP}\ }\textbf {\bibinfo {volume} {06}},\
  \bibinfo {pages} {004} (\bibinfo {year} {2003}{\natexlab{b}})},\ \Eprint
  {http://arxiv.org/abs/hep-ph/0303124} {arXiv:hep-ph/0303124} \BibitemShut
  {NoStop}%
%%CITATION = HEP-PH/0303124;%%
\bibitem [{\citenamefont {Shapiro}\ and\ \citenamefont
  {Sola}(2008)}]{Shapiro:2008yu}%
  \BibitemOpen
  \bibfield  {author} {\bibinfo {author} {\bibfnamefont {I.~L.}\ \bibnamefont
  {Shapiro}}\ and\ \bibinfo {author} {\bibfnamefont {J.}~\bibnamefont {Sola}},\
  }\href@noop {} {\  (\bibinfo {year} {2008})},\ \Eprint
  {http://arxiv.org/abs/0808.0315} {arXiv:0808.0315 [hep-th]} \BibitemShut
  {NoStop}%
\bibitem [{\citenamefont {Shapiro}\ and\ \citenamefont
  {Sol\`{a}}(2009)}]{Shapiro:2009dh}%
  \BibitemOpen
  \bibfield  {author} {\bibinfo {author} {\bibfnamefont {I.~L.}\ \bibnamefont
  {Shapiro}}\ and\ \bibinfo {author} {\bibfnamefont {J.}~\bibnamefont
  {Sol\`{a}}},\ }\href {\doibase 10.1016/j.physletb.2009.10.073} {\bibfield
  {journal} {\bibinfo  {journal} {Phys. Lett. B}\ }\textbf {\bibinfo {volume}
  {682}},\ \bibinfo {pages} {105} (\bibinfo {year} {2009})},\ \Eprint
  {http://arxiv.org/abs/0910.4925} {arXiv:0910.4925 [hep-th]} \BibitemShut
  {NoStop}%
%%CITATION = ARXIV:0910.4925;%%
\bibitem [{\citenamefont {Donoghue}(1994)}]{Donoghue:1994dn}%
  \BibitemOpen
  \bibfield  {author} {\bibinfo {author} {\bibfnamefont {J.~F.}\ \bibnamefont
  {Donoghue}},\ }\href {\doibase 10.1103/PhysRevD.50.3874} {\bibfield
  {journal} {\bibinfo  {journal} {Phys. Rev. D}\ }\textbf {\bibinfo {volume}
  {50}},\ \bibinfo {pages} {3874} (\bibinfo {year} {1994})},\ \Eprint
  {http://arxiv.org/abs/gr-qc/9405057} {arXiv:gr-qc/9405057} \BibitemShut
  {NoStop}%
\bibitem [{\citenamefont {Reuter}(1998)}]{Reuter:1996cp}%
  \BibitemOpen
  \bibfield  {author} {\bibinfo {author} {\bibfnamefont {M.}~\bibnamefont
  {Reuter}},\ }\href {\doibase 10.1103/PhysRevD.57.971} {\bibfield  {journal}
  {\bibinfo  {journal} {Phys.Rev.}\ }\textbf {\bibinfo {volume} {D57}},\
  \bibinfo {pages} {971} (\bibinfo {year} {1998})},\ \Eprint
  {http://arxiv.org/abs/hep-th/9605030} {arXiv:hep-th/9605030 [hep-th]}
  \BibitemShut {NoStop}%
%%CITATION = HEP-TH/9605030;%%
\bibitem [{\citenamefont {Bonanno}\ and\ \citenamefont
  {Reuter}(2000)}]{Bonanno:2000ep}%
  \BibitemOpen
  \bibfield  {author} {\bibinfo {author} {\bibfnamefont {A.}~\bibnamefont
  {Bonanno}}\ and\ \bibinfo {author} {\bibfnamefont {M.}~\bibnamefont
  {Reuter}},\ }\href {\doibase 10.1103/PhysRevD.62.043008} {\bibfield
  {journal} {\bibinfo  {journal} {Phys.Rev.}\ }\textbf {\bibinfo {volume}
  {D62}},\ \bibinfo {pages} {043008} (\bibinfo {year} {2000})},\ \Eprint
  {http://arxiv.org/abs/hep-th/0002196} {arXiv:hep-th/0002196} \BibitemShut
  {NoStop}%
%%CITATION = HEP-TH/0002196;%%
\bibitem [{\citenamefont {Giacchini}\ \emph {et~al.}(2020)\citenamefont
  {Giacchini}, \citenamefont {de~Paula~Netto},\ and\ \citenamefont
  {Shapiro}}]{Giacchini:2020zrl}%
  \BibitemOpen
  \bibfield  {author} {\bibinfo {author} {\bibfnamefont {B.~L.}\ \bibnamefont
  {Giacchini}}, \bibinfo {author} {\bibfnamefont {T.}~\bibnamefont
  {de~Paula~Netto}}, \ and\ \bibinfo {author} {\bibfnamefont {I.~L.}\
  \bibnamefont {Shapiro}},\ }\href {\doibase 10.1007/JHEP10(2020)011}
  {\bibfield  {journal} {\bibinfo  {journal} {JHEP}\ }\textbf {\bibinfo
  {volume} {10}},\ \bibinfo {pages} {011} (\bibinfo {year} {2020})},\ \Eprint
  {http://arxiv.org/abs/2009.04122} {arXiv:2009.04122 [hep-th]} \BibitemShut
  {NoStop}%
\bibitem [{\citenamefont {Reuter}\ and\ \citenamefont
  {Weyer}(2004)}]{Reuter:2003ca}%
  \BibitemOpen
  \bibfield  {author} {\bibinfo {author} {\bibfnamefont {M.}~\bibnamefont
  {Reuter}}\ and\ \bibinfo {author} {\bibfnamefont {H.}~\bibnamefont {Weyer}},\
  }\href {\doibase 10.1103/PhysRevD.69.104022} {\bibfield  {journal} {\bibinfo
  {journal} {Phys. Rev. D}\ }\textbf {\bibinfo {volume} {69}},\ \bibinfo
  {pages} {104022} (\bibinfo {year} {2004})},\ \Eprint
  {http://arxiv.org/abs/hep-th/0311196} {arXiv:hep-th/0311196} \BibitemShut
  {NoStop}%
%%CITATION = HEP-TH/0311196;%%
\bibitem [{\citenamefont {Shapiro}\ \emph {et~al.}(2005)\citenamefont
  {Shapiro}, \citenamefont {Sol\`{a}},\ and\ \citenamefont
  {Stefancic}}]{Shapiro:2004ch}%
  \BibitemOpen
  \bibfield  {author} {\bibinfo {author} {\bibfnamefont {I.~L.}\ \bibnamefont
  {Shapiro}}, \bibinfo {author} {\bibfnamefont {J.}~\bibnamefont {Sol\`{a}}}, \
  and\ \bibinfo {author} {\bibfnamefont {H.}~\bibnamefont {Stefancic}},\ }\href
  {\doibase 10.1088/1475-7516/2005/01/012} {\bibfield  {journal} {\bibinfo
  {journal} {JCAP}\ }\textbf {\bibinfo {volume} {01}},\ \bibinfo {pages} {012}
  (\bibinfo {year} {2005})},\ \Eprint {http://arxiv.org/abs/hep-ph/0410095}
  {arXiv:hep-ph/0410095} \BibitemShut {NoStop}%
%%CITATION = HEP-PH/0410095;%%
\bibitem [{\citenamefont {Babic}\ \emph {et~al.}(2005)\citenamefont {Babic},
  \citenamefont {Guberina}, \citenamefont {Horvat},\ and\ \citenamefont
  {Stefancic}}]{Babic:2004ev}%
  \BibitemOpen
  \bibfield  {author} {\bibinfo {author} {\bibfnamefont {A.}~\bibnamefont
  {Babic}}, \bibinfo {author} {\bibfnamefont {B.}~\bibnamefont {Guberina}},
  \bibinfo {author} {\bibfnamefont {R.}~\bibnamefont {Horvat}}, \ and\ \bibinfo
  {author} {\bibfnamefont {H.}~\bibnamefont {Stefancic}},\ }\href {\doibase
  10.1103/PhysRevD.71.124041} {\bibfield  {journal} {\bibinfo  {journal} {Phys.
  Rev. D}\ }\textbf {\bibinfo {volume} {71}},\ \bibinfo {pages} {124041}
  (\bibinfo {year} {2005})},\ \Eprint {http://arxiv.org/abs/astro-ph/0407572}
  {arXiv:astro-ph/0407572} \BibitemShut {NoStop}%
%%CITATION = ASTRO-PH/0407572;%%
\bibitem [{\citenamefont {Domazet}\ and\ \citenamefont
  {Stefancic}(2011)}]{Domazet:2010bk}%
  \BibitemOpen
  \bibfield  {author} {\bibinfo {author} {\bibfnamefont {S.}~\bibnamefont
  {Domazet}}\ and\ \bibinfo {author} {\bibfnamefont {H.}~\bibnamefont
  {Stefancic}},\ }\href {\doibase 10.1016/j.physletb.2011.07.038} {\bibfield
  {journal} {\bibinfo  {journal} {Phys. Lett. B}\ }\textbf {\bibinfo {volume}
  {703}},\ \bibinfo {pages} {1} (\bibinfo {year} {2011})},\ \Eprint
  {http://arxiv.org/abs/1010.3585} {arXiv:1010.3585 [gr-qc]} \BibitemShut
  {NoStop}%
%%CITATION = 1010.3585;%%
\bibitem [{\citenamefont {Koch}\ \emph {et~al.}(2015)\citenamefont {Koch},
  \citenamefont {Rioseco},\ and\ \citenamefont {Contreras}}]{Koch:2014joa}%
  \BibitemOpen
  \bibfield  {author} {\bibinfo {author} {\bibfnamefont {B.}~\bibnamefont
  {Koch}}, \bibinfo {author} {\bibfnamefont {P.}~\bibnamefont {Rioseco}}, \
  and\ \bibinfo {author} {\bibfnamefont {C.}~\bibnamefont {Contreras}},\ }\href
  {\doibase 10.1103/PhysRevD.91.025009} {\bibfield  {journal} {\bibinfo
  {journal} {Phys. Rev.}\ }\textbf {\bibinfo {volume} {D91}},\ \bibinfo {pages}
  {025009} (\bibinfo {year} {2015})},\ \Eprint {http://arxiv.org/abs/1409.4443}
  {arXiv:1409.4443 [hep-th]} \BibitemShut {NoStop}%
%%CITATION = ARXIV:1409.4443;%%
\bibitem [{\citenamefont {Bertini}(2025)}]{Bertini:2024naa}%
  \BibitemOpen
  \bibfield  {author} {\bibinfo {author} {\bibfnamefont {N.~R.}\ \bibnamefont
  {Bertini}},\ }\href {\doibase 10.1088/1475-7516/2025/04/071} {\bibfield
  {journal} {\bibinfo  {journal} {JCAP}\ }\textbf {\bibinfo {volume} {04}},\
  \bibinfo {pages} {071} (\bibinfo {year} {2025})},\ \Eprint
  {http://arxiv.org/abs/2408.05893} {arXiv:2408.05893 [gr-qc]} \BibitemShut
  {NoStop}%
\bibitem [{\citenamefont {Poisson}\ and\ \citenamefont
  {Will}(2014)}]{poisson2014gravity}%
  \BibitemOpen
  \bibfield  {author} {\bibinfo {author} {\bibfnamefont {E.}~\bibnamefont
  {Poisson}}\ and\ \bibinfo {author} {\bibfnamefont {C.~M.}\ \bibnamefont
  {Will}},\ }\href@noop {} {\emph {\bibinfo {title} {Gravity: Newtonian,
  post-newtonian, relativistic}}}\ (\bibinfo  {publisher} {Cambridge University
  Press},\ \bibinfo {year} {2014})\BibitemShut {NoStop}%
\bibitem [{\citenamefont {Will}(2018)}]{Will_2018}%
  \BibitemOpen
  \bibfield  {author} {\bibinfo {author} {\bibfnamefont {C.~M.}\ \bibnamefont
  {Will}},\ }\href@noop {} {\emph {\bibinfo {title} {Theory and Experiment in
  Gravitational Physics}}},\ \bibinfo {edition} {2nd}\ ed.\ (\bibinfo
  {publisher} {Cambridge University Press},\ \bibinfo {year}
  {2018})\BibitemShut {NoStop}%
\bibitem [{\citenamefont {Toniato}\ \emph {et~al.}(2017)\citenamefont
  {Toniato}, \citenamefont {Rodrigues}, \citenamefont {de~Almeida},\ and\
  \citenamefont {Bertini}}]{Toniato:2017wmk}%
  \BibitemOpen
  \bibfield  {author} {\bibinfo {author} {\bibfnamefont {J.~D.}\ \bibnamefont
  {Toniato}}, \bibinfo {author} {\bibfnamefont {D.~C.}\ \bibnamefont
  {Rodrigues}}, \bibinfo {author} {\bibfnamefont {A.~O.~F.}\ \bibnamefont
  {de~Almeida}}, \ and\ \bibinfo {author} {\bibfnamefont {N.}~\bibnamefont
  {Bertini}},\ }\href {\doibase 10.1103/PhysRevD.96.064034} {\bibfield
  {journal} {\bibinfo  {journal} {Phys. Rev. D}\ }\textbf {\bibinfo {volume}
  {96}},\ \bibinfo {pages} {064034} (\bibinfo {year} {2017})},\ \Eprint
  {http://arxiv.org/abs/1706.09032} {arXiv:1706.09032 [gr-qc]} \BibitemShut
  {NoStop}%
\bibitem [{\citenamefont {Sereno}\ and\ \citenamefont
  {Jetzer}(2006)}]{Sereno:2006re}%
  \BibitemOpen
  \bibfield  {author} {\bibinfo {author} {\bibfnamefont {M.}~\bibnamefont
  {Sereno}}\ and\ \bibinfo {author} {\bibfnamefont {P.}~\bibnamefont
  {Jetzer}},\ }\href {\doibase 10.1103/PhysRevD.73.063004} {\bibfield
  {journal} {\bibinfo  {journal} {Phys. Rev. D}\ }\textbf {\bibinfo {volume}
  {73}},\ \bibinfo {pages} {063004} (\bibinfo {year} {2006})},\ \Eprint
  {http://arxiv.org/abs/astro-ph/0602438} {arXiv:astro-ph/0602438} \BibitemShut
  {NoStop}%
%%CITATION = ASTRO-PH/0602438;%%
\bibitem [{\citenamefont {Toniato}\ \emph {et~al.}(2020)\citenamefont
  {Toniato}, \citenamefont {Rodrigues},\ and\ \citenamefont
  {Wojnar}}]{Toniato:2019rrd}%
  \BibitemOpen
  \bibfield  {author} {\bibinfo {author} {\bibfnamefont {J.~D.}\ \bibnamefont
  {Toniato}}, \bibinfo {author} {\bibfnamefont {D.~C.}\ \bibnamefont
  {Rodrigues}}, \ and\ \bibinfo {author} {\bibfnamefont {A.}~\bibnamefont
  {Wojnar}},\ }\href {\doibase 10.1103/PhysRevD.101.064050} {\bibfield
  {journal} {\bibinfo  {journal} {Phys. Rev. D}\ }\textbf {\bibinfo {volume}
  {101}},\ \bibinfo {pages} {064050} (\bibinfo {year} {2020})},\ \Eprint
  {http://arxiv.org/abs/1912.12234} {arXiv:1912.12234 [gr-qc]} \BibitemShut
  {NoStop}%
\bibitem [{\citenamefont {Bertini}\ and\ \citenamefont
  {Velten}(2023)}]{Bertini:2023pmp}%
  \BibitemOpen
  \bibfield  {author} {\bibinfo {author} {\bibfnamefont {N.~R.}\ \bibnamefont
  {Bertini}}\ and\ \bibinfo {author} {\bibfnamefont {H.}~\bibnamefont
  {Velten}},\ }\href {\doibase 10.1103/PhysRevD.107.124005} {\bibfield
  {journal} {\bibinfo  {journal} {Phys. Rev. D}\ }\textbf {\bibinfo {volume}
  {107}},\ \bibinfo {pages} {124005} (\bibinfo {year} {2023})},\ \Eprint
  {http://arxiv.org/abs/2303.09699} {arXiv:2303.09699 [gr-qc]} \BibitemShut
  {NoStop}%
\bibitem [{\citenamefont {Farina}\ \emph {et~al.}(2011)\citenamefont {Farina},
  \citenamefont {Kort-Kamp}, \citenamefont {Mauro},\ and\ \citenamefont
  {Shapiro}}]{Farina:2011me}%
  \BibitemOpen
  \bibfield  {author} {\bibinfo {author} {\bibfnamefont {C.}~\bibnamefont
  {Farina}}, \bibinfo {author} {\bibfnamefont {W.~J.~M.}\ \bibnamefont
  {Kort-Kamp}}, \bibinfo {author} {\bibfnamefont {S.}~\bibnamefont {Mauro}}, \
  and\ \bibinfo {author} {\bibfnamefont {I.~L.}\ \bibnamefont {Shapiro}},\
  }\href {\doibase 10.1103/PhysRevD.83.124037} {\bibfield  {journal} {\bibinfo
  {journal} {Phys. Rev. D}\ }\textbf {\bibinfo {volume} {83}},\ \bibinfo
  {pages} {124037} (\bibinfo {year} {2011})},\ \Eprint
  {http://arxiv.org/abs/1101.5611} {arXiv:1101.5611 [gr-qc]} \BibitemShut
  {NoStop}%
%%CITATION = ARXIV:1101.5611;%%
\bibitem [{\citenamefont {Rodrigues}\ \emph {et~al.}(2015)\citenamefont
  {Rodrigues}, \citenamefont {Chauvineau},\ and\ \citenamefont
  {Piattella}}]{Rodrigues:2015hba}%
  \BibitemOpen
  \bibfield  {author} {\bibinfo {author} {\bibfnamefont {D.~C.}\ \bibnamefont
  {Rodrigues}}, \bibinfo {author} {\bibfnamefont {B.}~\bibnamefont
  {Chauvineau}}, \ and\ \bibinfo {author} {\bibfnamefont {O.~F.}\ \bibnamefont
  {Piattella}},\ }\href {\doibase 10.1088/1475-7516/2015/09/009} {\bibfield
  {journal} {\bibinfo  {journal} {JCAP}\ }\textbf {\bibinfo {volume} {09}},\
  \bibinfo {pages} {009} (\bibinfo {year} {2015})},\ \Eprint
  {http://arxiv.org/abs/1504.05119} {arXiv:1504.05119 [gr-qc]} \BibitemShut
  {NoStop}%
\bibitem [{\citenamefont {Partanen}\ and\ \citenamefont
  {Tulkki}(2025)}]{Partanen:2025whp}%
  \BibitemOpen
  \bibfield  {author} {\bibinfo {author} {\bibfnamefont {M.}~\bibnamefont
  {Partanen}}\ and\ \bibinfo {author} {\bibfnamefont {J.}~\bibnamefont
  {Tulkki}},\ }\href@noop {} {\  (\bibinfo {year} {2025})},\ \Eprint
  {http://arxiv.org/abs/2506.14447} {arXiv:2506.14447 [gr-qc]} \BibitemShut
  {NoStop}%
\bibitem [{\citenamefont {Sultana}\ \emph {et~al.}(2012)\citenamefont
  {Sultana}, \citenamefont {Kazanas},\ and\ \citenamefont
  {Levi~Said}}]{Sultana:2012qp}%
  \BibitemOpen
  \bibfield  {author} {\bibinfo {author} {\bibfnamefont {J.}~\bibnamefont
  {Sultana}}, \bibinfo {author} {\bibfnamefont {D.}~\bibnamefont {Kazanas}}, \
  and\ \bibinfo {author} {\bibfnamefont {J.}~\bibnamefont {Levi~Said}},\ }\href
  {\doibase 10.1103/PhysRevD.86.084008} {\bibfield  {journal} {\bibinfo
  {journal} {Phys. Rev. D}\ }\textbf {\bibinfo {volume} {86}},\ \bibinfo
  {pages} {084008} (\bibinfo {year} {2012})},\ \Eprint
  {http://arxiv.org/abs/1910.06118} {arXiv:1910.06118 [gr-qc]} \BibitemShut
  {NoStop}%
\bibitem [{\citenamefont {Will}(2025)}]{Will:2025hwd}%
  \BibitemOpen
  \bibfield  {author} {\bibinfo {author} {\bibfnamefont {C.~M.}\ \bibnamefont
  {Will}},\ }\href {\doibase 10.1103/PhysRevD.111.104034} {\bibfield  {journal}
  {\bibinfo  {journal} {Phys. Rev. D}\ }\textbf {\bibinfo {volume} {111}},\
  \bibinfo {pages} {104034} (\bibinfo {year} {2025})},\ \Eprint
  {http://arxiv.org/abs/2503.03189} {arXiv:2503.03189 [gr-qc]} \BibitemShut
  {NoStop}%
\bibitem [{\citenamefont {{Lichnerowicz}}(1967)}]{1967rhm..book.....L}%
  \BibitemOpen
  \bibfield  {author} {\bibinfo {author} {\bibfnamefont {A.}~\bibnamefont
  {{Lichnerowicz}}},\ }\href@noop {} {\emph {\bibinfo {title} {{Relativistic
  Hydrodynamics and Magnetohydrodynamics}}}}\ (\bibinfo  {publisher} {W.A.
  Benjamin},\ \bibinfo {year} {1967})\BibitemShut {NoStop}%
\bibitem [{\citenamefont {Landau}\ and\ \citenamefont
  {Lifshitz}(1987)}]{landau_fluid_mechanics}%
  \BibitemOpen
  \bibfield  {author} {\bibinfo {author} {\bibfnamefont {L.~D.}\ \bibnamefont
  {Landau}}\ and\ \bibinfo {author} {\bibfnamefont {E.~M.}\ \bibnamefont
  {Lifshitz}},\ }\href@noop {} {\emph {\bibinfo {title} {Fluid Mechanics}}},\
  \bibinfo {edition} {2nd}\ ed.,\ \bibinfo {series} {Course of Theoretical
  Physics}, Vol.~\bibinfo {volume} {6}\ (\bibinfo  {publisher} {Pergamon
  Press},\ \bibinfo {year} {1987})\BibitemShut {NoStop}%
\bibitem [{\citenamefont {Anile}(1990)}]{Anile_1990}%
  \BibitemOpen
  \bibfield  {author} {\bibinfo {author} {\bibfnamefont {A.~M.}\ \bibnamefont
  {Anile}},\ }\enquote {\bibinfo {title} {Mathematical structure},}\ in\
  \href@noop {} {\emph {\bibinfo {booktitle} {Relativistic Fluids and
  Magneto-fluids: With Applications in Astrophysics and Plasma Physics}}},\
  \bibinfo {series and number} {Cambridge Monographs on Mathematical Physics}\
  (\bibinfo  {publisher} {Cambridge University Press},\ \bibinfo {year}
  {1990})\ p.\ \bibinfo {pages} {4–56}\BibitemShut {NoStop}%
\end{thebibliography}

\end{document}